\documentclass[reprint,
showpacs,preprintnumbers,
amsmath,amssymb,
aps,
pra]{revtex4-2}
\usepackage[dvipsnames]{xcolor}
\usepackage{graphicx}
\graphicspath{ {./Figures/Mollow} }

\usepackage{braket}
\usepackage{amsmath}
\usepackage{hyperref}

\usepackage[T1]{fontenc}

\DeclareUnicodeCharacter{2212}{-}

\begin{document}


\title{Quantum-correlated photons from spectrally-separated modes of a cavity coupled to a strongly-driven two-level atom}

\author{Alex Elliott}
\author{Jacob Ngaha}
\author{Scott Parkins}
\affiliation{%
Te Whai Ao--Dodd-Walls Centre for Photonic and Quantum Technologies}
\affiliation{%
Department of Physics, University of Auckland, Auckland 1010, New Zealand.
}

\author{Takao Aoki}
\affiliation{%
Department of Applied Physics, Waseda University, 3-4-1 Okubo, Shinjuku-ku, Tokyo 169-8555, Japan
}

\date{\today}

\begin{abstract}
Photon counting statistics are explored, theoretically, from a pair of cavity modes coupled to the fluorescent transitions in a strongly-driven two-level atom. We show that the cavity modes acquire nonclassical photon statistics that are representative of dressed-state picture atomic transitions. In particular, the modes are shown to be antibunched, while simultaneously having a cross-correlation value greater than unity. Furthermore, we propose an implementation of the system with a nanofiber cavity QED system, based on a strongly-driven cesium atom.

\end{abstract}

\pacs{Valid PACS appear here}
\maketitle


\section{Introduction}
Interactions between light and two-level systems have been studied since the very early days of quantum optics. Despite their simplicity, they exhibit a remarkable range of physical phenomena. In the regime of weak coherent driving, the need for an atomic re-excitation process imposes a temporal separation between successively emitted photons, hence the antibunched photon counting statistics of resonance fluorescence \cite{Carmichael1976,Kimble1977}. However, a strong incident field dresses the bare states into a ladder of doublets; the emission spectrum splits into a three-peaked structure known as the Mollow triplet \cite{Mollow1969, Mollow1975, Wu1975}. In the dressed-state interpretation, the three peaks arise from distinct transitions between eigenstates of the quantized resonance fluorescence Hamiltonian, which have both atomic and photonic characteristics \cite{Cohen_Tannoudji1977}. 

The persistence of whole-field antibunching in the strong-driving limit is explained as an interference effect between different transition pathways \cite{Dalibard1983}. Isolating specific components of the emitted field unveils properties of the dressed state transitions that manifest themselves as frequency-dependent photon correlations \cite{Arnoldus1984}. The correlation structure can be explored by spectrally filtering the emitted field, and shows potential for developing strongly-correlated photon sources \cite{Ulhaq2012, Masters2023, Ngaha2024}. Furthermore, the nuanced correlation structure has shown promise for the generation of photonic Bell states \cite{Liu2024, Hu2025}.

Beyond free-space resonance fluorescence, the interaction can be enhanced and modified by coupling to a near-resonant cavity mode. In the Purcell regime, this can enhance the atomic emission rate whilst retaining the spectral structure, enabling rapid production of photonic Bell states \cite{Wang2025}. Coupling a specific dressed-state transition to a cavity mode has been shown to affect the width and relative prominence of the emission peaks, but the triplet structure is still retained \cite{Zhou1998,Kim2014}. However, the simultaneous coupling to an independent pair of cavity modes, one centered on each sideband of the Mollow triplet, has not previously been studied. This is partly due to resonators tending to rely on an extremely short length to elicit a strong atom-light coupling, giving a correspondingly large free spectral range (FSR). On the contrary, nanofiber-based cavities are typically much longer, with the strong coupling arising from extreme transverse field confinement in the tapered nanofiber section \cite{Kato2015,Schneeweiss2016}. These cavities therefore can have a smaller FSR \cite{Nemet2020}, comparable in scale to the sideband splittings that have been demonstrated with an alkali atom \cite{Ng2022}. Conceivably, nanofiber cavities can be engineered such that an adjacent pair of frequency-resolved cavity modes are resonant at each of the Mollow fluorescence frequencies, and thereby simultaneously interact with the same atom. One then might expect the output from each of these modes to be representative of, or related to, the photon statistics for the relevant peaks of the Mollow spectrum. In this paper, we consider exactly this scenario: coupling a strongly-driven two-level atom to a pair of cavity modes, to explore the photon correlations that arise.

The paper is organized as follows. First, in Section~\ref{sect2}, we briefly introduce the Mollow spectrum, describe the atom-cavity system, and discuss the calculation of statistical properties of the fields. In Section~\ref{Two-Level Model}, we consider an idealized model of a strongly-driven two-level atom interacting with an adjacent pair of cavity modes, to unveil the expected photon statistics for the respective cavity modes and understand the dependence on the interaction parameters. In Section~\ref{Alkali Atom Model}, we describe and numerically integrate a model of a \textsuperscript{133}Cs atom, coherently driven and coupled to a pair of cavity modes on the cycling transition. This allows us to predict the viability of the scheme with a real atom, using parameters relevant to recent nanofiber cavity QED experiments. We finish with some concluding remarks in Section~\ref{Conclusion}.

\section{Single-Atom Cavity QED System}
\label{sect2}
When an atom is illuminated by a coherent laser field, the scattered light contains an incoherent component that grows increasingly prominent with laser intensity. When the Rabi frequency $\Omega$ is sufficiently large, the power spectrum of scattered light acquires a characteristic triplet structure, centered on the bare atomic resonance \cite{Mollow1969}. The Mollow triplet can be interpreted in a basis of dressed eigenstates of the resonance fluorescence Hamiltonian \cite{Cohen_Tannoudji1977}. Each manifold of $N$ total excitations is split into a doublet, with corresponding energies that are separated by the Rabi frequency. Three distinct transition energies are resonant between respective pairs in each adjacent manifold; a central peak at the bare atomic resonance, and a pair of fluorescence peaks respectively blue- and red-shifted by $\pm\Omega$.

\begin{figure}[htp]
	\centering
    \includegraphics{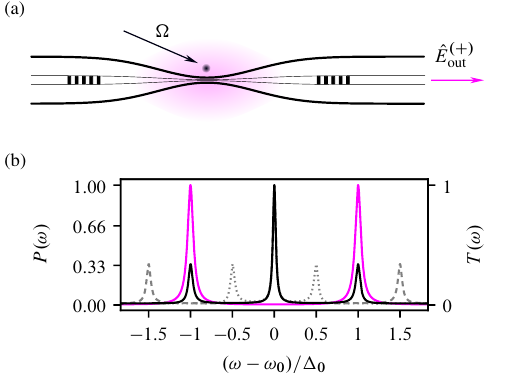}
	\caption{(a) Schematic of a single-atom in the evanescent field of a nanofiber cavity. The atom is driven by an auxiliary laser with Rabi frequency $\Omega$. The bare atomic frequency is precisely midway between the adjacent cavity resonances, which are separated by $2\Delta_0$. (b) Representative transmission spectrum, $T(\omega)$, for horizontally-polarized modes of a long cavity, showing two adjacent resonances (solid magenta), and atomic power spectrum, $P(\omega)$, for a two-level atom driven with Rabi frequencies $\Omega = \{0.5,~1,~1.5\}\Delta_0$ (dotted, solid, and dashed black, respectively), where $\omega_0$ is the bare atomic resonance.}
	\label{TransmissionSpectrum}
\end{figure}

The emission from resonance fluorescence is randomly scattered into free space, and cannot generally be captured entirely. Instead, a portion of the field is typically channeled with a high numerical aperture lens \cite{Ng2022} or by Purcell enhancement with a broadband cavity \cite{Wang2025}. Properties of the captured field can then be investigated, and supplemented with further filtering to isolate specific frequencies. We propose a novel set-up that siphons specific frequency components of the triplet directly into fiber, by coupling the atom simultaneously to adjacent modes of a nanofiber cavity, as depicted in Fig.~\ref{TransmissionSpectrum}(a). The configuration of cavity resonances that we consider is depicted in Fig.~\ref{TransmissionSpectrum}(b); two adjacent cavity modes are tuned to be resonant half an FSR above and below the bare atomic transition frequency, respectively. If the atom is driven with a Rabi frequency equal to half the FSR, that is $\Omega = \Delta_0$, the fluorescence sidebands of the Mollow triplet will each be resonant with one of the cavity modes.

\subsection{Characterizing the Fields}
The steady-state power spectrum of a source is calculated from two-time correlations in the output field, through the Fourier transform
\begin{equation}
	P(\omega) = \frac{1}{2\pi}\int_{-\infty}^{\infty}\braket{\hat{O}^\dagger(0),\hat{O}(t')}_\text{ss}e^{-i\omega t'}dt',
\end{equation}
where $\hat{O}$ is an output field operator, and $\braket{\hat{O}^\dagger,\hat{O}}_\text{ss} \equiv \braket{\hat{O}^\dagger\hat{O}}_\text{ss} -|\braket{\hat{O}}_\text{ss}|^2$, with the subscripts indicating steady-state expectations. In particular, the atomic emission spectrum is calculated using the operator $\sqrt{\gamma}\hat{\sigma}_-$, where $\hat\sigma_-$ is the Pauli lowering operator and $\gamma$ is the atomic linewidth (FWHM). Emission spectra for each mode of the cavity are calculated using the operators $\sqrt{2\kappa}\hat{r}$ and $\sqrt{2\kappa}\hat{b}$, respectively, where $\hat{r}$ and $\hat{b}$ are annihilation operators for red- and blue-detuned cavity modes, and $\kappa$ is the cavity field decay rate. The spectrum of the total cavity field (restricted to the two modes of interest) is calculated using the operator $\hat{E}_\text{out}^{(+)}=\sqrt{2\kappa}\hat{E}^{(+)}$, where $\hat{E}^{(+)} = \hat{r} +\hat{b}$ (and $\hat{E}^{(-)} = \hat{r}^\dagger +\hat{b}^\dagger$). Two-time correlations are calculated numerically, with the quantum regression formulae, using the relevant master equation \cite{Carmichael1999,qutip5}.

The nature of photon correlations are quantified by the normalized correlation functions
\begin{equation}
    g^{(2)}_{\hat{O}_1\hat{O}_2}(\tau) = \frac{\braket{\hat{O}^\dagger_1(0)\hat{O}^\dagger_2(\tau)\hat{O}_2(\tau)\hat{O}_1(0)}_\text{ss}}{\braket{\hat{O}_1^\dagger\hat{O}_1}_\text{ss}\braket{\hat{O}_2^\dagger\hat{O}_2}_\text{ss}},
\end{equation}
where the numerator is again calculated with the quantum regression formulae. Note that for auto-correlations (i.e., $\hat{O}_1=\hat{O}_2$) the subscript operator is labeled only once. The second-order correlations for a classical field are subject to Cauchy-Schwarz inequalities. For a single mode, these classical limits are, 
\begin{equation}
    g_{\hat{O}}^{(2)}(0)\geq1,~~~g_{\hat{O}}^{(2)}(0)\geq g_{\hat{O}}^{(2)}(\tau).
    \label{ineq11}
\end{equation}
For two modes of a field, there is also a classical limit on the extent of correlation between modes, relative to the auto-correlations within them \cite{Wasak2014},
\begin{equation}
	|g^{(2)}_{\hat{O}_1\hat{O}_2}(\tau)|\leq \sqrt{g^{(2)}_{\hat{O}_1}(0)g^{(2)}_{\hat{O}_2}(0)}.
	\label{ineq12}
\end{equation} 
We consider and compare the spectrum and second-order correlations for both an idealized two-level atom, and an implementation with a cesium atom, in the following sections.

\section{Two-Level Model}
\label{Two-Level Model}
\subsection{Multimode Jaynes-Cummings Model}
We start with an idealized model of a two-level atom, coupled to a pair of cavity modes. If the atom is resonantly driven, the Jaynes-Cummings model extends to give the interaction picture Hamiltonian (with respect to the atomic transition frequency)
\begin{equation}
\begin{split}
	\hat{H} =& \frac{\Omega}{2}\left(\hat{\sigma}_-+ \hat{\sigma}_+\right)+\Delta_0\big(\hat{b}^\dagger\hat{b} -\hat{r}^\dagger\hat{r}\big)\\
    &+  g\left(\big[\hat{r}+\hat{b}\big]\hat{\sigma}_+ + \hat{\sigma}_-\big[\hat{b}^\dagger + \hat{r}^\dagger\big]\right),
    \end{split}
    \label{Hammy}
\end{equation}
where the atom-cavity coupling strength $g$ is assumed to be real and the identical for both cavity modes. 

The composite density operator, $\hat{\rho}$ for the system evolves according to the Lindblad master equation ($\hbar = 1$),
\begin{equation}
    \dot{\hat{\rho}} = -i\big[\hat{H},\hat{\rho}\big] + \gamma\mathcal{D}(\hat{\sigma}_-)[\hat{\rho}]+ 2\kappa\mathcal{D}(\hat{r})[\hat{\rho}]+ 2\kappa\mathcal{D}(\hat{b})[\hat{\rho}],
\end{equation}
with the dissipative superoperators $\mathcal{D}(\hat{O})[\hat{\rho}] = \hat{O}\hat{\rho}\hat{O}^\dagger - \frac{1}{2}\hat{O}^\dagger\hat{O}\hat{\rho}- \frac{1}{2}\hat{\rho}\hat{O}^\dagger\hat{O}$. This model assumes that each cavity mode has the same decay rate, and furthermore that each decays independently, which is valid as long as they are well resolved in frequency, i.e., $\Delta_0\gg \{g,~\kappa\}$. To illustrate the operating principles of the scheme, we focus on the specific case of $\Delta_0 = 25\gamma$.

\begin{figure}[t]
	\centering
	\includegraphics{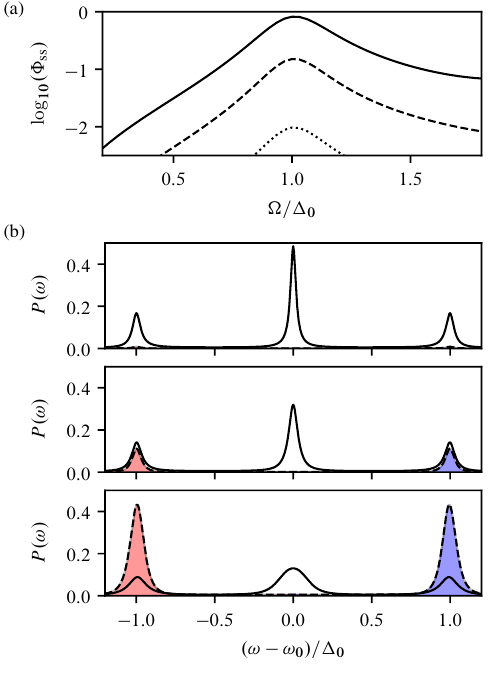}
	\caption{(a) Log of total steady-state cavity flux, $\Phi_\text{ss}=2\kappa\braket{\hat{E}^{(-)}\hat{E}^{(+)}}_\text{ss}$ as a function of laser Rabi frequency, for cavities with a halfwidth $\kappa = 2.5\gamma$, and atom-cavity coupling strengths of $g = \{0.25,1,2.5\}\gamma$ for the dotted, dashed, and solid lines respectively. In all cases the flux is maximum when the resonance condition $\Omega=\Delta_0$ is satisfied. (b) Atomic (solid lines), and cavity (dashed line) power spectra  corresponding to the parameters shown in (a), i.e. $g = \{0.25,1,2.5\}\gamma$ from top to bottom. Also shown in shaded red and blue are the spectra of each respective cavity mode. As the atom-cavity coupling strength is increased, the free-space atomic emission is suppressed because emission is preferentially channeled through the cavity. Furthermore, the relative prominence of the elastic peak is diminished due to the the cavity enhancement of the fluorescent transitions.}
	\label{fig2}
\end{figure}

Fig.~\ref{fig2}(a) shows the cavity flux as a function of Rabi frequency, for a few values of the atom-cavity coupling strength $g$. The flux is maximum when the Rabi frequency is chosen appropriately for the respective cavity FSR (i.e., $\Omega = \Delta_0$), so that the Mollow sidebands are each resonant with a cavity mode. Choosing the driving strengths accordingly, Fig.~\ref{fig2}(b) shows the atomic and cavity power spectra for the corresponding cavity configurations. In each case, the overall cavity emission is doubly peaked, representing the two populated modes. In the limit of weak coupling, the cavity essentially acts as a pair of band-pass filters that siphon a portion of the atomic emission spectrum without significantly affecting the structure. As the coupling strength increases, the widths and relative prominence of the atomic emission peaks are modified. The central peak is diminished as cavity modes enhance the fluorescent sideband transitions, thereby diminishing resonant scattering. Furthermore, the peaks are broadened by cavity modes because of the modified vacuum fluctuations in the vicinity of the sidebands \cite{Lewenstein1987}.

\subsection{Dressed-State Model}
A dressed-state model for the interaction can provide valuable insight into the effective dynamics. We define dressed states in terms of the bare atomic states,
\begin{equation}
    \ket{\pm} = \frac{1}{\sqrt{2}}\left(\ket{g}\pm\ket{e}\right),
\end{equation}
and define dressed state operators on this basis,
\begin{equation}
    \hat{\sigma}_z^D = \ket{+}\bra{+}-\ket{-}\bra{-},~~~\hat{\sigma}_\pm^D = \ket{\pm}\bra{\mp}.
\end{equation}
By choosing an appropriate interaction picture, and making the secular approximation, the Hamiltonian (\ref{Hammy}) can be expressed in the atomic dressed-state basis as
\begin{equation}
	\hat{H} =\frac{g}{2}\bigg[\Big( \hat{r}-\hat{b}^\dagger\Big)\hat{\sigma}_-^D +\hat{\sigma}_+^D  \Big(\hat{r}^\dagger-\hat{b} \Big)\bigg],
\end{equation}
as described in Appendix~\ref{appendix}. From this Hamiltonian, and incorporating cavity loss, one can derive Heisenberg-Langevin equations of motion for the cavity mode annihilation operators (also in Appendix~\ref{appendix}). These equations can be formally integrated, and for $\kappa\gg g,\gamma$, one obtains solutions of the form
\begin{align}
    \hat{r}(t) &\simeq \frac{-ig}{2\kappa}\hat{\sigma}_+^D(t) +  \sqrt{2\kappa}\int_{-\infty}^t e^{-\kappa(t-t')}\hat{r}_{\text{in}}(t')dt',
    \\
    \hat{b}(t) &\simeq \frac{ig}{2\kappa}\hat{\sigma}_-^D(t)+  \sqrt{2\kappa}\int_{-\infty}^t e^{-\kappa(t-t')}\hat{b}_{\text{in}}(t')dt' ,
\end{align}
where $\hat{r}_{\text{in}}(t)$ and $\hat{b}_{\text{in}}(t)$ are input quantum noise operators that satisfy the commutation relations $[\hat{r}_{\text{in}}(t),\hat{r}^\dagger_{\text{in}}(t')] = [\hat{b}_{\text{in}}(t),\hat{b}^\dagger_{\text{in}}(t')] = \delta(t-t')$.  
Assuming that the cavity input fields are vacuum and therefore do not contribute to photon correlations, one can see from the above equations that, in this limit of weak coupling, the cavity correlations are representative of the underlying, filtered atomic dressed-state correlations (see Appendix~\ref{appendix} for more discussion).

\subsection{Photon Correlations}
\begin{figure}[t]
	\centering
	\includegraphics{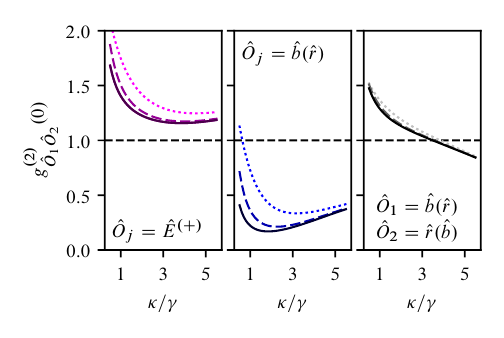}
	\caption{Initial value of the auto-correlation for the total cavity output field (left) and for an individual cavity field mode (center), and cross-correlation between cavity modes (right), as the cavity decay rate is varied, and with $\Omega=\Delta_0 = 25\gamma$. Each line shows a different value of the atom-cavity coupling strength, with $g/\gamma = \{0.25,~1,~2.5\}$ for the solid, dashed, and dotted curves respectively.}
	\label{fig3}
\end{figure}

Second-order coincidence correlations for the cavity emission are shown in Fig.~\ref{fig3}. The total emitted cavity field ($\hat E^{(+)}$) is bunched for the parameters that we consider, from simultaneous scattering in each of the two fluorescence sidebands \cite{Masters2023}. Nonetheless, each of the separate cavity modes is antibunched because of the inherently discrete nature of the fluorescent dressed state transitions. When the cavity bandwidth becomes too small, the antibunching is lost due to the correlations being dominated by interference effects of the cavities, rather than the underlying source statistics \cite{Nienhuis1993, Joosten2000}. The cross-correlation between the pair of cavity modes shows bunching due to the correlated emission of light in the red and blue fluorescence transitions, so long as the cavity resonance is sharp enough to effectively isolate the components. If the cavity bandwidth is too large, photons from the elastic scattering peak can also couple into the cavity modes, which commensurately affects the nature of the photon correlations.

For example, when both cavities have parameters $g=\kappa=\gamma$, the overall output field is bunched, with $g^{(2)}_{\hat{E}^{(+)}}(0) = 1.5$. The auto-correlation for each cavity mode shows antibunching, with $g^{(2)}_{\hat{b}}(0) = 0.36$, while the cross-correlation shows bunching between the modes as $g^{(2)}_{\hat{b}\hat{r}}(0) = 1.32$.  For all values shown in Fig.~\ref{fig3}, the correlations clearly violate both the single- and two-mode Cauchy-Schwarz inequalities Eqs.~(\ref{ineq11}) and (\ref{ineq12}) (with $\tau=0$).

\begin{figure}[b]
	\centering
	\includegraphics{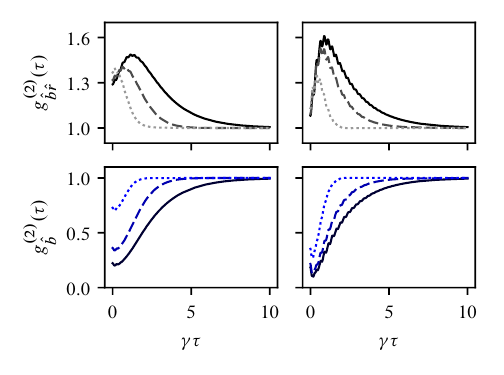}
	\caption{Second-order cross-correlations (top row) and auto-correlation (bottom row) for two different cavity halfwidths, $\kappa = \gamma$ (left) and $\kappa = 2.5\gamma$ (right). The lines show the correlation for different atom-cavity coupling strengths, with $g = \{0.25,~1,~2.5\}\gamma$ for the solid, dashed, and dotted curves respectively.}
	\label{fig4}
\end{figure}

Another key aspect of the photon statistics, beyond coincidence, is the time dependence of correlations. Fig.~\ref{fig4} shows the time-dependent photon correlations from the cavity modes, for some sample cavity parameters. It is evident that the emission violates the single-mode time-dependent Cauchy-Schwarz inequalities Eqs.~(\ref{ineq11}) and (\ref{ineq12}) over timescales of a few atomic lifetimes. When the cavity decay rate is appropriately chosen, and the coupling weak, the cross-correlations take an initial value close to unity. This is caused by an interference effect that arises when filtering the Mollow spectrum side-peaks \cite{Schrama1992}. However, for higher cooperativity cavities, the coupling allows the two modes to be initially more strongly correlated. This increased cross-correlation comes with the consequence of reducing the anti-correlation within the modes.

\section{Alkali Atom Model}
\label{Alkali Atom Model}
\begin{figure}[htp]
	\centering
    \includegraphics{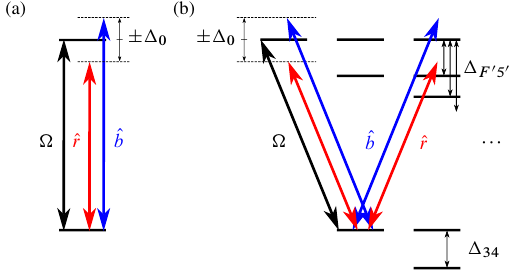}
	\caption{Schematic of (a) the two-level atom model and (b) an implementation within the cycling transition of \textsuperscript{133}Cs, showing some of the relevant D2 line hyperfine sublevels.}
	\label{CsSchematicTLA2cav}
\end{figure}

Having examined an idealized model of the system, we now model a potential implementation of the scheme in an alkali-atom-based, nanofiber cavity QED system. In order to realize two-level behavior within the multilevel cesium atom, the D2-line cycling transition (i.e., $\ket{6S_{1/2},F=4,m_F = \pm4}\leftrightarrow\ket{6P_{3/2},F=5,m_F = \pm5}$) is driven with a circularly-polarized ($q=\pm1$) laser. Nanofiber cavities generally support linearly-polarized eigenmodes, so an adjacent pair of the horizontally-polarized modes are used to couple to the closed transition. Birefringence of the fiber shifts the vertically-polarized modes away from the relevant frequencies, and they can therefore be neglected. The specific configuration we consider is shown in Fig.~\ref{CsSchematicTLA2cav}, illustrating transitions involving the targeted $m_F = -4$ ground state. Cavity-induced transitions to the $m_{F'}=-3'$ states are suppressed as long as $|\Delta_0-\Delta_{F'5'}|\gg\gamma$ for each of the hyperfine excited states, where $\gamma/2\pi = 5.2\,\text{MHz}$. This is helped by the favorable Clebsch-Gordan coefficients in cesium; the ratio of coupling strength on the cycling transition compared to the $m_{F'}=-3'$ states is $0.15,~0.34,~\text{and}~0.44$, for the $F'=5', 4'~\text{and}~3'$ states, respectively \cite{SteckCs133}. 

The system is described in an interaction picture on a basis of the D2-line hyperfine sublevels, with the ground (excited) states labeled as $\ket{F,m_F}$ ($\ket{F',m_{F'}}$). The interaction-picture ground and excited state energies are specified relative the $F=4$ and $F'=5'$ states, respectively, giving the bare atomic Hamiltonian
\begin{equation}
\begin{split}
    \hat{H}_0 = &\sum_{m_F}\Delta_{34}\ket{3,m_F}\bra{3,m_F}\\
    &+\sum_{F'}\sum_{m_{F'}}\Delta_{F'5'}\ket{F',m_{F'}}\bra{F',m_{F'}}.
\end{split}
\end{equation}
Polarization-dependent dipole transition operators are defined to describe coupling between the ground and excited states,
\begin{equation}
    \hat{D}_q = \sum_{F,F'}\sum_{m_F}C(F,F',m_F,q)\ket{F,m_F}\bra{F',m_F+q},
\end{equation}
where $C(F,F',m_F,q)$ is a Clebsch-Gordan coefficient, and $q=0$ or $q=\pm1$ for $\pi$- and $\sigma_\pm$-polarized transitions, respectively. The coefficients are normalized to unity on the cycling transition, i.e., $C(4,5',\pm 4,\pm1)=1$. In this notation, horizontally-polarized transitions are expressed as a sum of circular polarizations, $\hat{D}_{\rm h} = \frac{1}{\sqrt{2}}(\hat{D}_+ + \hat{D}_-)$. The electric dipole coupling Hamiltonian is then expressed compactly as 
\begin{equation}
	\begin{split}
	\hat{H} &= \hat{H}_0 + \Delta_0\left(\hat{b}^\dagger\hat{b}-\hat{r}^\dagger\hat{r}\right)+\frac{\Omega}{2}\left(\hat{D}_{-}+ \hat{D}_{-}^\dagger\right) \\
		&~~~+g_{\rm h}\left(\Big[\hat{r} + \hat{b}\Big]\hat{D}^\dagger_{\rm h} + \hat{D}_{\rm h}\Big[\hat{r}^\dagger + \hat{b}^\dagger\Big]\right),
\end{split}
\end{equation}
where introducing $g_{\rm h} = \sqrt{2}g$ compensates for the scaling factor in the horizontal dipole operator, for the sake of comparison with the two-level-atom case. The cesium atom master equation has three distinct terms for atomic decay, to describe spontaneous emission for each polarization, as 
    \begin{equation}
    \dot{\hat{\rho}} = -i\big[\hat{H},\hat{\rho}\big] + \gamma\sum_q\mathcal{D}(\hat{D}_q)[\hat{\rho}]+ 2\kappa\mathcal{D}(\hat{r})[\hat{\rho}]+ 2\kappa\mathcal{D}(\hat{b})[\hat{\rho}].
\end{equation}
That is, spontaneous emission is treated collectively with respect to the hyperfine levels, and independently with respect to polarization. As in the simple model, we focus on the special case of $\Delta_0 = \Omega = 25\gamma$.

\begin{figure}[htp]
	\centering
	\includegraphics{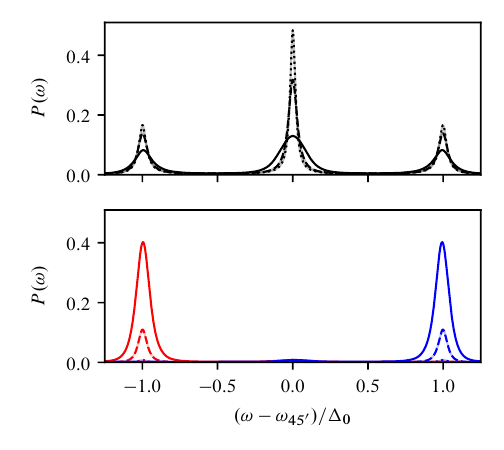}
	\caption{(Top) Cesium atomic emission spectra and (bottom) red and blue cavity power spectra, calculated from the D2-line interaction using the cesium hyperfine structure model, where $\omega_{45'}$ is the bare $F=4\leftrightarrow F'=5'$ transition frequency. The cavity modes both have a halfwidth $\kappa = 2.5\gamma$, and an atom-cavity coupling strength of $g = \{0.25,1,2.5\}\gamma$, for the dotted, dashed, and solid lines, respectively. }
	\label{fig6}
\end{figure}

Fig.~\ref{fig6} displays the steady-state spectra of atomic and cavity emission from the cesium system, showing equivalent features to those predicted by the simple model. In the regimes we consider, the atom remains well confined to the cycling transition; the most extreme parameters still retain $>97\%$ population confinement. Additional repumping lasers and/or external magnetic field Zeeman shifts can enhance the confinement if desired, but we do not explicitly implement these methods here.

\begin{figure}[htp]
	\centering
	\includegraphics{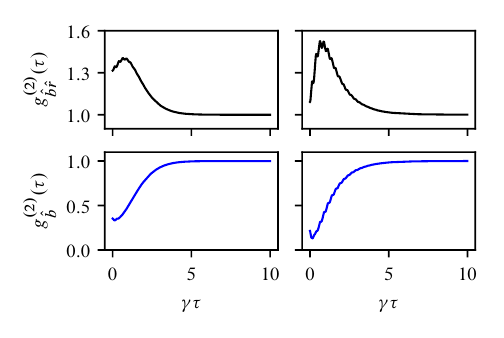}
	\caption{(Top) Cross-correlations and (bottom) auto-correlations for the cavity modes obtained from the cesium hyperfine structure model. In both cases $g=\gamma$, while (left) $\kappa = \gamma$ and (right) $\kappa = 2.5\gamma$.}
	\label{fig7}
\end{figure}

With the model configured appropriately, Fig.~\ref{fig7} shows the second-order auto- and cross-correlations for the cavity modes coupled to the cesium atom. The system exhibits nonclassical features that are almost identical to those predicted in the ideal two-level model. The cavity modes display antibunched photon counting statistics, and cross-correlations that surpass the initial auto-correlation value, in violation of the Cauchy-Schwarz inequalities, Eqs.~(\ref{ineq11}) and (\ref{ineq12}). In particular, for $\kappa=g=\gamma$, the modes have an autocorrelation of $g^{(2)}_{\hat{b}}(0) = 0.35$ and an initial cross-correlation $g^{(2)}_{\hat{b}\hat{r}}(0) = 1.32$. For $\kappa=2.5g=2.5\gamma$, the modes are more strongly antibunched, with $g^{(2)}_{\hat{b}}(0) = 0.21$, and show an initial cross-correlation of $g^{(2)}_{\hat{b}\hat{r}}(0) = 1.09$. This system therefore shows promise as a way to generate both one-mode and two-mode nonclassical light, from a single atomic transition, and directly into fiber.

\section{Conclusion}
\label{Conclusion}
We have proposed a method to generate nonclassical photon statistics in the output field of a pair of cavity modes, coupled to a single two-level atom. This builds on previous work that examined the correlations between different frequencies of the Mollow spectrum. By working with a long cavity that has a relatively small FSR, two adjacent modes can couple to the same atomic transition, transferring the atomic correlations to the pair of cavity modes. The spectral-filtering and cavity enhancement that is inherent to this scheme could prove useful for the generation of entangled photon pairs, and photonic Bell states \cite{Liu2024,Wang2025,Hu2025}. We also demonstrated that implementing this method should be viable with a real atom-cavity system, using a model of a single \textsuperscript{133}Cs atom, and with parameters relevant to a contemporary nanofiber cavity QED system. 

\section*{Acknowledgments} 
AE thanks Quantum Technologies Aotearoa (QTA) for supporting his visit to the group of TA at Waseda University as part of this research. SP also thanks the group of TA for support and hospitality during a visit to Waseda University.

\bibliography{References}
\appendix

\onecolumngrid

\section{Atomic Dressed-State Picture}
\label{appendix}
\noindent
Bare-state atomic raising ($\hat{\sigma}_+ = \ket{e}\bra{g}$) and lowering ($\hat{\sigma}_- = \ket{g}\bra{e}$) operators
are first expressed in terms of the dressed basis states, 
\begin{equation}
    \ket{\pm} = \frac{1}{\sqrt{2}}\left(\ket{g}\pm\bra{e}\right).
\end{equation}
Introducing dressed-state spin operators 
\begin{equation}
    \hat{\sigma}_z^D = \ket{+}\bra{+}-\ket{-}\bra{-},~~~\hat{\sigma}_-^D = \ket{-}\bra{+},~~~\hat{\sigma}_+^D = \ket{+}\bra{-},
\end{equation}
the Pauli raising and lowering operators can be expressed as 
\begin{equation}
    \hat{\sigma}_\pm = \frac{1}{2}\hat{\sigma}_z^D \mp \frac{1}{2}\Big(\hat{\sigma}_+^D-\hat{\sigma}_-^D\Big).
\end{equation}
Using these expressions, the Hamiltonian (\ref{Hammy}) becomes
\begin{equation}
    \hat{H} = \Delta_0\Big(\hat{b}^\dagger\hat{b} - \hat{r}^\dagger\hat{r}\Big) + \frac{\Omega}{2}\hat{\sigma}_z^D + g\bigg[ \Big(\hat{r}+\hat{b}\Big)\left[\frac{1}{2}\hat{\sigma}_z^D - \frac{1}{2}\Big(\hat{\sigma}_+^D-\hat{\sigma}_-^D\Big)\right] + \left[\frac{1}{2}\hat{\sigma}_z^D + \frac{1}{2}\Big(\hat{\sigma}_+^D-\hat{\sigma}_-^D\Big)\right]\Big(\hat{r}^\dagger+\hat{b}^\dagger\Big) \bigg].
\end{equation}
Moving into a new interaction picture defined by the transformation Hamiltonian
\begin{equation}
    \hat{H}_0 = \Delta_0\Big(\hat{b}^\dagger\hat{b} - \hat{r}^\dagger\hat{r}\Big) + \frac{\Omega}{2}\hat{\sigma}_z^D ,
\end{equation}
the operators pick up the following explicit time dependencies
\begin{equation}
    \hat{b}\rightarrow\hat{b}e^{-i\Delta_0 t},~~~\hat{r}\rightarrow\hat{r}e^{i\Delta_0 t},~~~\hat{\sigma}_-^D\rightarrow\hat{\sigma}_-^De^{-i\Omega t},~~~\hat{\sigma}_+^D\rightarrow\hat{\sigma}_+^De^{i\Omega t}.
\end{equation}
A secular approximation neglects terms that oscillate  like $e^{\pm i\Delta_0 t}$ or $e^{\pm i(\Omega+\Delta_0)t}$, on the grounds that $\Omega,\Delta_0\gg g$. With the resonance condition $\Omega = \Delta_0$, one arrives at
\begin{equation}
    \hat{H}_\text{int} =\frac{g}{2}\bigg[\Big( \hat{r}-\hat{b}^\dagger\Big)\hat{\sigma}_-^D +\hat{\sigma}_+^D  \Big(\hat{r}^\dagger-\hat{b} \Big)\bigg].
\end{equation}
With this effective Hamiltonian, and taking into account cavity field-mode decay at rate $\kappa$, we can write down Heisenberg-Langevin equations for the cavity field operators in the forms
\begin{equation}
    \begin{split}
        \dot{\hat{r}}&=\frac{-ig}{2}\hat{\sigma}_+^D -\kappa\hat{r} +\sqrt{2\kappa}\,\hat{r}_{\text{in}}(t),\\
        \dot{\hat{b}}&=\frac{ig}{2}\hat{\sigma}_-^D -\kappa\hat{b} +\sqrt{2\kappa}\,\hat{b}_{\text{in}}(t) ,
    \end{split}
\end{equation}
where $\hat{r}_{\text{in}}(t)$ and $\hat{b}_{\text{in}}(t)$ are input-field quantum noise operators satisfying the standard commutation relations $[\hat{r}_{\text{in}}(t),\hat{r}^\dagger_{\text{in}}(t')] =[\hat{b}_{\text{in}}(t),\hat{b}^\dagger_{\text{in}}(t')] =  \delta(t-t')$ \cite{Gardiner1985}. We also assume that $\hat{r}_{\text{in}}(t)$ and $\hat{b}_{\text{in}}(t)$ are independent, i.e., $[\hat{r}_{\text{in}}(t),\hat{b}_{\text{in}}(t')] =[\hat{r}_{\text{in}}(t),\hat{b}^\dagger_{\text{in}}(t')]= 0$. Steady-state solutions for the operators are found by formally integrating the above equations:
\begin{equation}
\begin{split}
    \hat{r}(t) &= \frac{-ig}{2}\int_{-\infty}^t e^{-\kappa(t-t')}\hat{\sigma}_+^D(t')dt' +  \sqrt{2\kappa}\int_{-\infty}^t e^{-\kappa(t-t')}\hat{r}_{\text{in}}(t')dt',\\
    \hat{b}(t) &= \frac{ig}{2}\int_{-\infty}^t e^{-\kappa(t-t')}\hat{\sigma}_-^D(t')dt' +  \sqrt{2\kappa}\int_{-\infty}^t e^{-\kappa(t-t')}\hat{b}_{\text{in}}(t')dt'.
\end{split}
\end{equation}
In the case that $\kappa$ is much larger than $g$ and $\gamma$, then these solutions are well approximated by
\begin{equation}
    \hat{r}(t) \simeq \frac{-ig}{2\kappa}\hat{\sigma}_+^D(t) +  \sqrt{2\kappa}\int_{-\infty}^t e^{-\kappa(t-t')}\hat{r}_{\text{in}}(t')dt',~~~\hat{b}(t) \simeq \frac{ig}{2\kappa}\hat{\sigma}_-^D(t)+  \sqrt{2\kappa}\int_{-\infty}^t e^{-\kappa(t-t')}\hat{b}_{\text{in}}(t')dt'.
\end{equation}
Assuming vacuum field inputs to the two cavity modes, it follows that the red- and blue-detuned cavity modes exhibit photon statistics that are representative of the underlying filtered dressed state transitions. In particular, from \cite{Schrama1992,Ngaha2024}, the auto-correlations are of the approximate forms
\begin{equation}
    g^{(2)}_{\hat{r}}(\tau)\simeq 1-e^{-\frac{\gamma}{2}\tau},~~~g^{(2)}_{\hat{r}}(\tau)\approx1-e^{-\frac{\gamma}{2}\tau},
\end{equation}
while the cross-correlations are expressed as 
\begin{equation}
    g^{(2)}_{\hat{r}\hat{b}}(\tau)\simeq e^{-\frac{\gamma}{2}\tau}-1+\frac{1}{2}\big(2-e^{-\kappa\tau}\big)^2 + \frac{1}{2}e^{-2\kappa\tau}.
\end{equation}
These expressions predict an initial autocorrelation value of $g^{(2)}_{\hat{r}}(0)\simeq 0$ ($g^{(2)}_{\hat{b}}(0)\simeq 0$), and an initial cross-correlation $g^{(2)}_{\hat{r}\hat{b}}(0)\simeq 1$, and show reasonably good agreement with the results shown in Figs.~\ref{fig4} and \ref{fig7} for the largest value of $\kappa$ considered and with the smaller values of $g$.

\end{document}